\documentclass[a4paper]{jpconf}
\usepackage{graphicx}
\begin{document}
\title{Search for Neutrinoless Double Beta Decay with NEMO~3 and SuperNEMO}

\author{Stefan S\"oldner-Rembold on behalf of the NEMO~3 and SuperNEMO Collaborations}

\address{University of Manchester, School of Physics and Astronomy, Oxford Road, Manchester, M13 9PL, UK}

\ead{stefan.soldner-rembold@manchester.ac.uk}

\begin{abstract}
Since 2003 the NEMO~3 experiment has been searching for neutrinoless double beta decay using about 10~kg of enriched 
isotopes.
A limit of $T_{1/2}^{0\nu}>5.8 \times 10^{23}$~years at $90\%$ CL
has been obtained for $^{100}$Mo from the first two years of data. 
Several  measurements of $2\nu\beta\beta$ decays have also been performed.
A first NEMO~3 measurement of the $2\nu\beta\beta$ half-life of $^{130}$Te is presented, giving a
value of $T_{1/2}^{2\nu}=(7.6 \pm 1.5 \mbox{(stat)} \pm 0.8 \mbox{(syst)})\times 10^{20}$~years. 
In parallel, there is an active R\&D programme for the SuperNEMO experiment which
is expected to commence data taking in 2012--2013 with 100--200~kg of enriched isotopes.
\end{abstract}

\vspace{-7mm}
\section{Introduction}
Neutrinoless double beta decay ($0\nu\beta\beta$)
leads to the decay of a nucleus of charge $Z$ and atomic number $A$ via the process $(A,Z) \to (A,Z+2) + 2e^-$.
This decay violates total lepton number $L$. Its observation would
therefore be a direct indication for physics beyond the Standard Model.
Neutrinos would be Majorana and not Dirac particles, thereby solving one
of the fundamental questions of particle physics. The $0\nu\beta\beta$ half-life is given by
\begin{equation}
\frac{1}{T^{0\nu}_{1/2}(A,Z)}=|M^{0\nu}(A,Z)|^2 G^{0\nu}(Q,Z) \langle m_{\beta\beta} \rangle ^2,
\label{eq1}
\end{equation}
where $M^{0\nu}(A,Z)$ is the nuclear matrix element (NME) and $G^{0\nu}(Q,Z)$ is a known phase
space factor that depends on the $Q$-value of the process. The half-life is proportional to the squared effective
Majorana mass, $\langle m_{\beta\beta} \rangle^2,$ which is given by a  sum over the masses $m_i$ of the mass eigenstates,
$\langle m_{\beta\beta} \rangle=\sum_{i=1,2,3} U_{ei}^2 m_i$, weighted by the squared elements, $U^2_{ei}$, of the PMNS neutrino
mixing matrix. The pairing term in the nuclear binding energy leads to a splitting of the parabola
describing the binding energy for isobaric nuclei with even $A$. Beta decay
is therefore forbidden or strongly suppressed for even-even nuclei, making them the only candidates 
for observing $0\nu\beta\beta$ decays.

\vspace{-4mm}
\section{NEMO-3}
\label{sec-nemo3}
The NEMO~3\footnote{Neutrino Ettore Majorana Observatory} experiment has been taking data since 2003 in
the Modane Underground Laboratory (LSM) located in the Frejus tunnel at a depth of 4800~m water equivalent. 
The experiment has a cylindrical shape with 20 sectors that contain different isotopes in the form
of thin foils with a total surface of about 20~m$^2$ (Table~\ref{tab1}).
The main isotopes used for the $0\nu\beta\beta$ search are about 7~kg of $^{100}$Mo and about 1~kg of  $^{82}$Se. Smaller
amounts of other isotopes are mainly used for measuring the $2\nu\beta\beta$ process. Tellurium
and copper foils are used for background measurements. 
\begin{table}[ht]
\caption{\label{tab1}Isotopes used in NEMO~3.}
\begin{center}
\begin{tabular}{l|c|c|c|c|c|c|c|c|c}
\br
 & $^{100}$Mo & $^{82}$Se & $^{116}$Cd & $^{96}$Zr & $^{150}$Nd & $^{48}$Ca & $^{130}$Te & $^{\rm nat}$Te & Cu \\
\mr 
mass [g] & 6914 & 932  & 405  & 9.4  & 37.0 & 7.0  & 454  & 491 & 621 \\
$Q$ [keV]& 3034 & 2995 & 2805 & 3350 & 3367 & 4772 & 2529 &     &      \\
\br
\end{tabular}
\end{center}
\end{table}
On each side of the foils is a $\sim 50$~cm wide tracking volume consisting of a total of 6180 drift cells operated in Geiger mode
with a typical vertex resolution of 5~mm and 8~mm in the coordinates transverse and perpendicular to the foil,
respectively.
The drift gas is helium with admixtures of $4\%$ ethyl alcohol, $1\%$ argon and $0.1\%$ water.
A $25$~Gauss magnetic field created by a solenoid provides charge identification. The calorimeters comprises 1940 plastic
scintillators coupled to low radioactivity photomultipliers. For 1~MeV electrons the energy resolution (FWHM) ranges
from $14.1\%$ to $17.7\%$ and the timing resolution is 250~ps. The timing
measurement is used to identify background sources originating from outside the foils (external background).
Identification of photons, electrons, positrons and alpha particles is a powerful tool to reject internal
background from the foil and other background from inside the detector volume. 
In the data analysis the kinematics of the different background sources are simulated using Monte Carlo generators. 
The rates are determined from HPGe measurements of the material and through measurements of control channels
such as the emission of an electron in association with a photon or an alpha particle.

A major background is due to the $\beta$ decay of $^{214}$Bi which has a high $Q$-value of 3.27 keV.
It is produced in the decay chain of radon outgassed by the rock surrounding the detector.
Radon purification has been obtained using charcoal filters trapping radon which subsequently
decays in the filter with a half-life of 3.8 days. The $^{214}$Bi contamination can be measured
using the BiPo process, where $^{214}$Bi decays into $^{214}$Po via $\beta$ decay and subsequently,
with a half-life of $164$~$\mu$s, into $^{210}$Pb. Measurements of this process demonstrate that
the radon contamination has been significantly decreased since the installation of the radon
filter in October 2004 (phase II). The data taken before October 2004 are labeled phase I.

\subsection{Measurement of $2\nu\beta\beta$ decays}
\begin{figure}[ht]
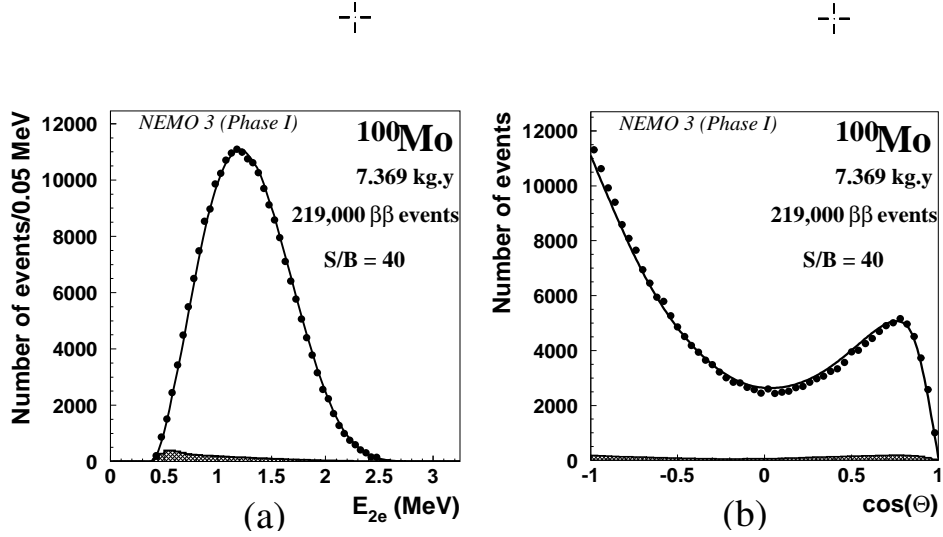

\begin{center}
\begin{minipage}{13pc}
\includegraphics[width=14pc]{figure2a.epsi}
\end{minipage}\hspace{2pc}%
\begin{minipage}{13pc}
\includegraphics[width=14pc]{figure2b.epsi}
\end{minipage}\hspace{2pc}%
\caption{\label{fig1} Distributions of the a) energy sum, $E_{12}$, and b)
angular variable, $\cos\theta$, of the two electrons observed in
$^{100}$Mo decays. The phase I data are shown as points, the subtracted background as hashed histogram.
The data are compared to $2\nu\beta\beta$ Monte Carlo.}
\end{center}
\end{figure}
The NEMO~3 experiment is in the unique position to perform high statistics measurements
of $2\nu\beta\beta$ decays, $(A,Z) \to (A,Z+2) + 2e^- + 2\nu$~\cite{nemo2}.\footnote{This process is not the same
as two subsequent $\beta$ decays, which would be energetically forbidden. It was first predicted in 
1935 by Goeppert-Mayer~\protect\cite{gm}.} These measurements improve the understanding of the $2\nu\beta\beta$
process which is the ultimate background in the $0\nu\beta\beta$ search. Its contribution can only be reduced
by improving the energy resolution. Furthermore, the measured $2\nu\beta\beta$ rates help to constrain nuclear
models and NME calculations which are currently a source of large uncertainty when translating the $0\nu\beta\beta$ half-lifes
into an effective Majorana neutrino mass, $\langle m_{\beta\beta} \rangle$ (Eq.~\ref{eq1}).
The distribution of the energy sum of the two electrons and their angular distribution are shown
in Fig.~\ref{fig1} for $^{100}$Mo~\cite{nemo1}. The agreement of the data with a $2\nu\beta\beta$ simulation is generally
good apart from a small shift in the angular distribution. The measured $^{100}$Mo half-life
is $T_{1/2}^{2\nu}=(7.11 \pm 0.02 \mbox{(stat)} \pm 0.54 \mbox{(syst)})\times 10^{18}$~years. 

The $2\nu\beta\beta$ half-life of $^{130}$Te has been a long-standing mystery due to the wide
range of measurements using geochemical sources. The half-life seems to depend on the age
of the sample used, $\simeq 7-9 \times 10^{20}$ years~\cite{young} for samples with an age of the order
100 million years and $\simeq 25-27 \times 10^{20}$ years~\cite{old} for samples older than 1 billion years. 
It has even been speculated that this could be related to time dependence
of the Fermi constant, $G_F$. A discussion of the measurements can be found in~\cite{bara}.
The first indication of a positive result in $^{130}$Te was obtained using TeO$_2$ crystals and yielded a value
$T_{1/2}^{2\nu}=(6.1 \pm 1.4 \mbox{(stat)} \pm^{2.9}_{3.5} \mbox{(syst)})\times 10^{18}$~years~\cite{arna}.
\begin{figure}[ht]
\begin{center}
\includegraphics[width=14pc]{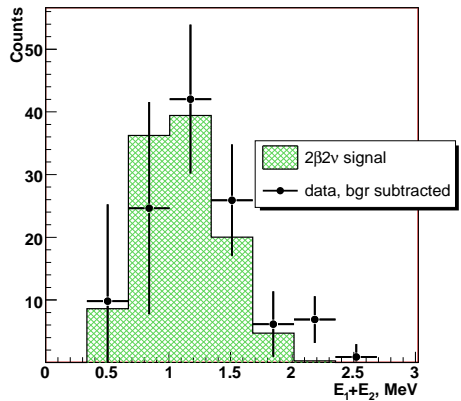}\hspace{2pc}%
\begin{minipage}[b]{13pc}
\caption{\label{te130} Distribution of the energy sum, $E_{12}$, of the two electrons observed
in $^{130}$Te decays after background subtraction (phase I + phase II, 534 days).
}
\end{minipage}
\end{center}
\end{figure}

A more accurate measurement of the $2\nu\beta\beta$ half-life has recently been performed by NEMO~3 using 
454~g of $^{130}$Te. The data, corresponding to 534 days, are shown in Fig.~\ref{te130} after
background subtraction. The background subtracted distribution contains $109 \pm 22$ events. This corresponds to
$T_{1/2}^{2\nu}=(7.6 \pm 1.5 \mbox{(stat)} \pm 0.8 \mbox{(syst)})\times 10^{20}$~years. The value
is consistent with the measurement of~\cite{arna} and with the lower values of the geochemical
experiments. An overview of all $2\nu\beta\beta$ half-lives measured by NEMO~3 is given in Table~\ref{tab2}.
\begin{table}[ht]
\caption{\label{tab2} Half-life of $2\nu\beta\beta$ measured using the phase I data taken by NEMO~3 (360 days).
The $^{130}$Te measurement uses phase I and phase II data (534 days)}
\begin{center}
\begin{tabular}{r|c|c}
\br
isotope & signal/background & $T_{1/2}$ [$10^{19}$ years] \\
\mr 
 $^{100}$Mo & $40$ & $0.711\pm0.002\mbox{ (stat) }\pm 0.054 \mbox{ (syst) }$\\
 $^{82}$Se  & $4$   & $9.6 \pm0.3\mbox{ (stat) }\pm 1.0 \mbox{ (syst) }$\\
 $^{116}$Cd & $7.5$   & $2.8 \pm0.1\mbox{ (stat) }\pm 0.3\mbox{ (syst) }$\\
 $^{150}$Nd  & $2.8$    & $0.97 \pm0.07\mbox{ (stat) }\pm 0.1\mbox{ (syst) }$\\
 $^{96}$Zr   & $1$   & $2.0 \pm0.3\mbox{ (stat) }\pm 0.2\mbox{ (syst) }$\\
 $^{48}$Ca  & $\sim 10$   & $3.9 \pm0.7\mbox{ (stat) }\pm 0.6\mbox{ (syst) }$\\
 $^{130}$Te & $0.25$  & $76 \pm 15 \mbox{ (stat) }\pm 8 \mbox{ (syst) }$\\
\br
\end{tabular}
\end{center}
\end{table}


\subsection{Search for Neutrinoless Double Beta Decay}
\begin{figure}[ht]
\begin{minipage}{11pc}
\includegraphics[width=12pc]{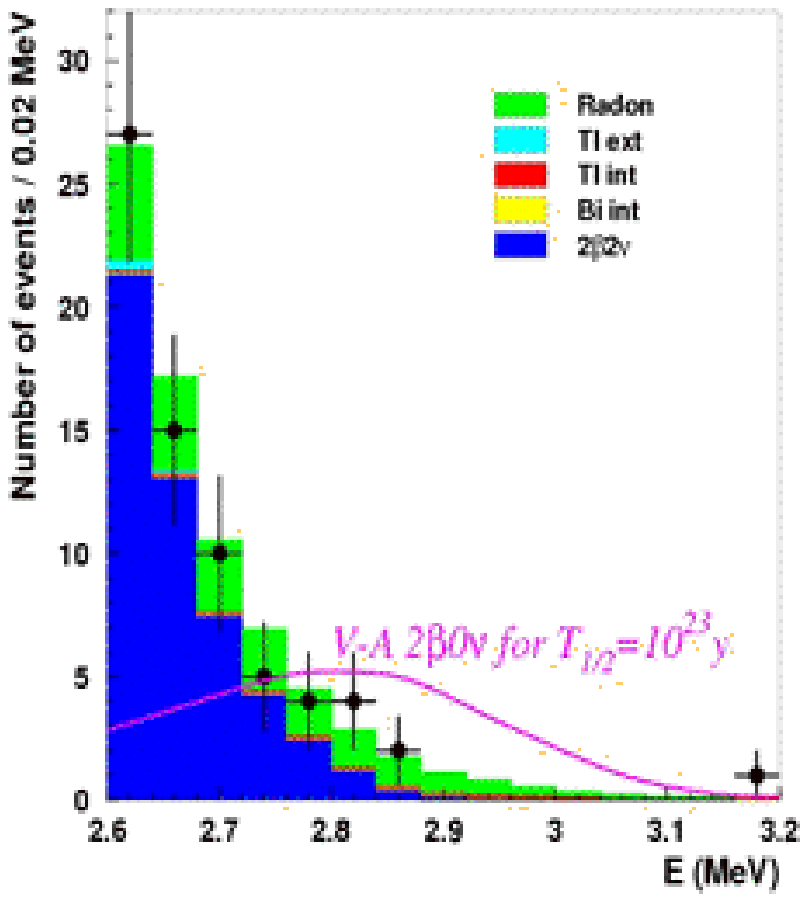}
\end{minipage}\hspace{2pc}%
\begin{minipage}{11pc}
\includegraphics[width=12pc]{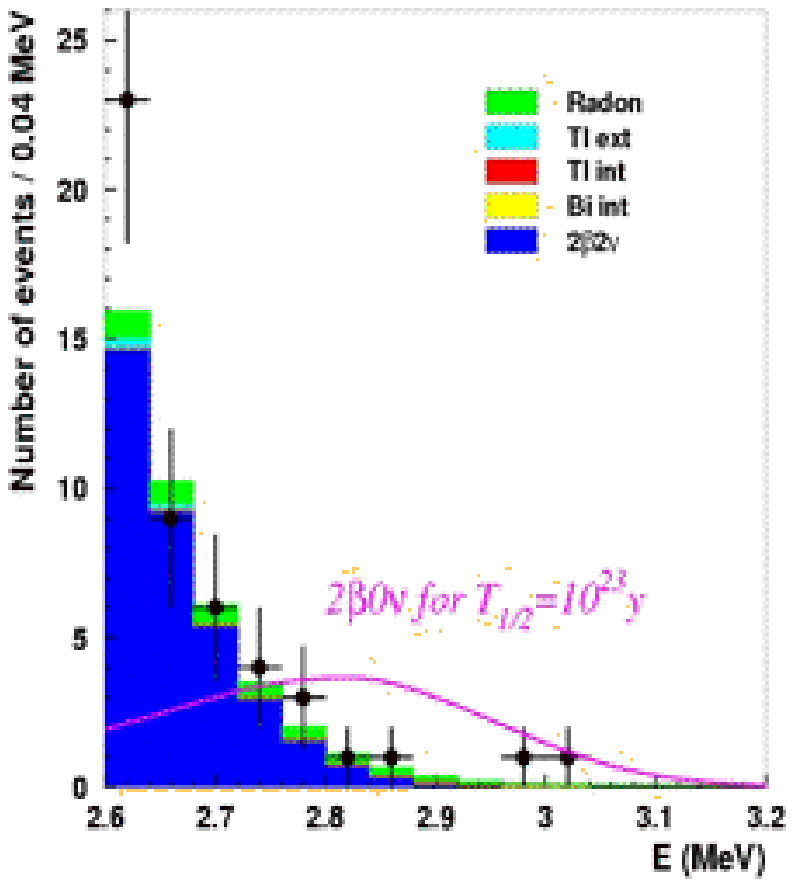}
\end{minipage}\hspace{2pc}%
\begin{minipage}{11pc}
\includegraphics[width=12pc]{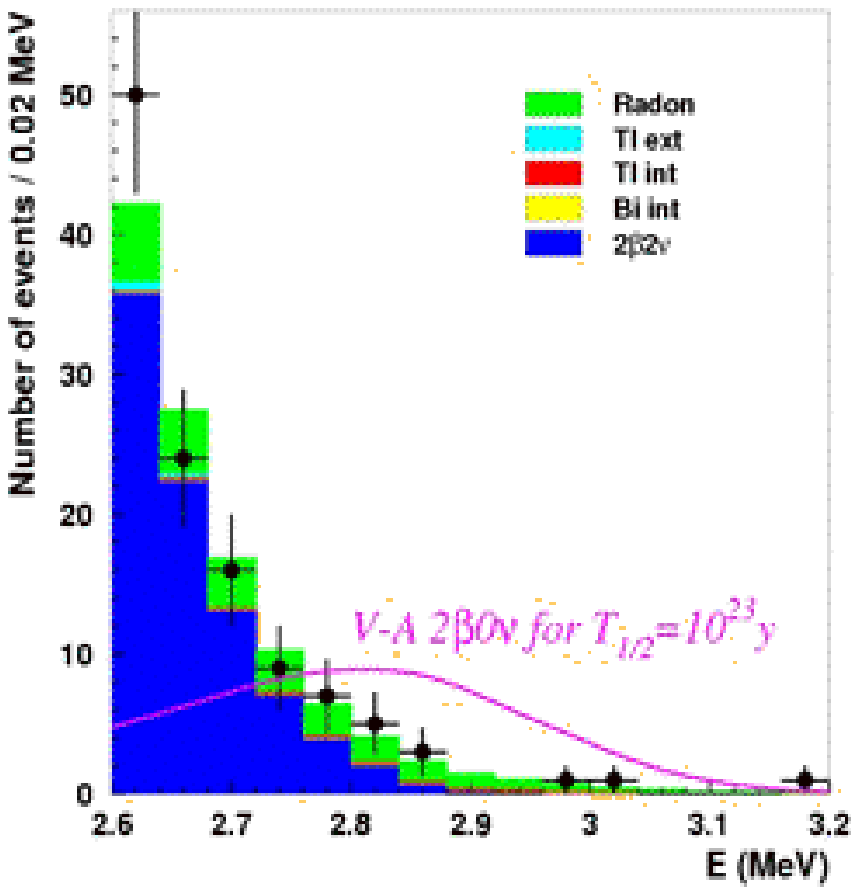}
\end{minipage}
\caption{\label{fig4} Distribution of the energy sum, $E_{12}$, of the two electrons for
$^{100}$Mo. The data are shown as points,  $2\nu\beta\beta$ background is shown in blue, the radon induced background
in green and the signal distribution in magenta. a) phase I (394 days, high radon); b) phase II (299 days, low radon); 
c) phase I+II (693 days).} 
\end{figure}

The distribution of the energy sums, $E_{12}$, of the two electrons is used to search for neutrinoless double beta decay.
A signal would correspond to an excess at $E_{12}\approx Q$, smeared out by the energy resolution of the calorimeter.
A Monte Carlo simulation of a signal is shown in Fig.~\ref{fig4} for $^{100}$Mo. 
The number of background events expected in the energy window $2.8<E_{12}<3.2$~MeV is 12.1 for the sum
of phase I and II, corresponding to 693 days of data taking, and the number of observed events is 11. 
Limits on the half-life are set using a maximum likelihood technique, yielding $T_{1/2}^{0\nu}>5.8 \times 10^{23}$~years
at $90\%$ Confidence Level (CL). Depending on the values of NME used~\cite{nme}, this translates into a limit
on the neutrino mass of $\langle m_{\beta\beta} \rangle < 0.8 - 1.3$~eV\@.
In 2006, the collaboration has decided to perform a blind analysis with the current data set and plans to update
the results in summer 2008 and again in early 2010.

Another possibility for neutrinoless double beta decays is Majoron emission,
$(A,Z) \to (A,Z+2) + 2e^- + \chi^0$~\cite{maj}. Majorons are Goldstone bosons arising due to the spontaneous breaking of
the global $B-L$ symmetry, where $B$ is the baryon number. Other possibilities for $0\nu\beta\beta$ decay arise
in supersymmetric models with $R$-parity violation with the emission of two $\chi^0$~\cite{moha}.
Limits are expressed in terms of the spectral index $n$, which is defined by the phase space of the emitted
particles, $G^{0\nu}\sim(Q-E_{12})^n$. The distribution of the energy sum is distorted depending on the spectral index $n$ 
and the limits are set using a maximum likelihood method. The limits are given in Table~\ref{tab-maj}.
For single Majoron emission, the half-life $T^{-1}_{1/2}$ is proportional to $|\langle g_{ee} \rangle|^2$, where
$g_{ee}$ is the Majoron-neutrino coupling constant. For the case $n=1$ limits of $\langle g_{ee} \rangle <
(0.4-1.9) \times 10^{-4}$ and $\langle g_{ee} \rangle <(0.66-1.7) \times 10^{-4}$ are obtained for $^{100}$Mo
and $^{82}$Se, respectively. Limits have also been set on an admixture of a right-handed current in
the Lagrangian.
\begin{table}[ht]
\caption{\label{tab-maj} Limits on the half-life at $90\%$ CL for models with Majoron emission, where $n$
is the spectral index (phase I data) and on a right-handed (V+A) contribution in the Lagrangian.}
\begin{center}
\begin{tabular}{c|c|c|c|c|c}
\br
isotope & $n=1$ & $n=2$ & $n=3$ & $n=7$ & V + A\\
\mr 
 $^{100}$Mo & $2.7\times 10^{22}$ y & $1.7\times 10^{22}$ y & $1.0\times 10^{22}$ y & $7\times 10^{19}$ y 
& $3.2\times 10^{23}$ y  \\
 $^{82}$Se  & $1.5\times 10^{22}$ y & $6.0\times 10^{21}$ y & $3.1\times 10^{21}$ y & $5.0\times 10^{20}$ y
& $1.2\times 10^{23}$ y  \\
\br
\end{tabular}
\end{center}
\end{table}

\vspace{-4mm}
\section{SuperNEMO}
The SuperNEMO experiment will be based on the successful NEMO~3 concept. The unique features
of this tracking plus calorimetry approach are:
\begin{itemize}
\item Measurement of process kinematics:

The measurement of the main kinematic observables, the individual electron 
energies and their angular
correlation will be used to 
study the underlying physics mechanism (e.g. SUSY, right-handed
currents) of the $0\nu\beta\beta$ process.

\item Sources separated from the detector:

This allows to measure several isotopes. This is essential 
to reduce systematic
uncertainties, to confirm a $0\nu\beta\beta$ discovery and to
identify the underlying physics mechanism.

\item Particle identification:

Electron, positron, gamma and alpha identification are powerful tools for background rejection.
Photon identification is used to reject any unknown nuclear gamma line.

\end{itemize}
The SuperNEMO Collaboration comprises about 60 physicists from 12 countries. Major R\&D projects
have been approved in France and the UK. The preliminary SuperNEMO design is based on a planar, modular geometry
with 20 modules each containing about 5~kg of enriched isotopes. The main challenges for the SuperNEMO
design are addressed in this R\&D project:
\begin{itemize}
\item
Improvement of the the calorimeter energy resolution to $4\%$ at electron energies
of $3$~MeV is necessary to discriminate $0\nu\beta\beta$ decays from background.
To accomplish this goal, several ongoing studies are investigating the choice of calorimeter
parameters such as scintillator material, the shape, size and coating of calorimeter blocks,
combined with low radioactivity photomultipliers with high quantum efficiency. Initial
studies have demonstrated excellent energy resolution ($\sim 6.5\%$ at 1 MeV) for small
size samples; the focus is now to retain this property in larger blocks.
\begin{table}[ht]
\caption{\label{tab3} A comparison of the main NEMO~3 and SuperNEMO parameters.}
\begin{center}
\begin{tabular}{c|c|c}
\br
  & NEMO~3 & SuperNEMO \\
\mr 
isotope & $^{100}$Mo & $^{150}$Nd or $^{82}$Se \\  
mass    & 7 kg & 100--200 kg \\
signal efficiency & $8\%$ & $>30\%$ \\
$^{208}$Tl in foil & $<20 \mu$Bq/kg  & $<2 \mu$Bq/kg \\
$^{214}$Bi in foil & $<300 \mu$Bq/kg  & $<10 \mu$Bq/kg ($^{82}$Se) \\
energy resolution & $8\%$ & $4\%$ \\
(FWHM at 3 MeV) & & \\
half-life  & $T_{1/2}^{0\nu}>2\times 10^{24}$ years & $T_{1/2}^{0\nu}>2\times 10^{26}$ years \\
neutrino mass & $\langle m_{\beta\beta} \rangle < 0.3 - 1.3$ eV & $\langle m_{\beta\beta} \rangle < 50 - 110$ meV \\
\br
\end{tabular}
\end{center}
\end{table}
\item The tracker will consist of Geiger cells similar to NEMO~3 with a length
of $\sim$ 4~m. Several smaller prototypes have been built and successfully operated 
to optimize the design
of the tracker, including the cell size, wire geometry and wire diameters.
Large prototypes ($\sim$~100--300 cells) will be constructed in the near future.
In parallel, a wiring robot is being designed and tested to allow for large scale production.
\item
The choice of isotope is based on a set of parameters: a long $2\nu\beta\beta$ half-life,
a high $Q$-value, a large phase space factor $G^{0\nu}$ and a large NME. The enrichment
possibility on a large scale is also a factor in selecting the isotope. The main
candidate isotopes have emerged to be $^{82}$Se and $^{150}$Nd. As sample of 4~kg
of $^{82}$Se has been enriched and is currently undergoing purification.
The collaboration is investigating the possibility of enriching large amounts
of $^{150}$Nd via the method of atomic laser isotope separation.
\item 
Ultra-low background levels are of paramount importance for the discovery potential of SuperNEMO.
The source foils must be pure and their radioactive contamination must be precisely
measured. The most important contaminants are $^{208}$Tl and $^{214}$Bi. To measure
their activities, dedicated detectors for the BiPo process (see Section~\ref{sec-nemo3}) are developed.
The first BiPo prototype was installed in the Canfranc laboratory in 2006.
\end{itemize}
The improvements expected when going from NEMO~3 to SuperNEMO are given in Table~\ref{tab3}.
Based on the results from the ongoing R\&D projects, including detailed
detector and physics simulations, a Technical Design Report (TDR) will be written in 2009.
First modules can be installed as early as 2010 and all twenty modules will be
running by 2012--2013.

\vspace{-4mm}
\section{Summary}
The NEMO approach of using tracking plus calorimetry is unique for a $0\nu\beta\beta$ experiment
and allows for excellent background rejection, choice of multiple isotopes and full
kinematic reconstruction. NEMO~3 is taking data, measuring the $2\nu\beta\beta$
process for several nuclei and setting limits on the half-life of the $0\nu\beta\beta$ process.
At the same there is an active R\&D programme for the next-generation experiment SuperNEMO.

\end{document}